\documentclass[reprint,
 amsmath,amssymb,
 aps,prd,
 superscriptaddress,
 twocolumn, longbibliography
]{revtex4-1}

\usepackage{graphicx}

\usepackage{dcolumn}
\usepackage{bm}
\usepackage{comment}
\usepackage{xspace}
\usepackage{units}
\usepackage{float}
\usepackage{multirow}
\usepackage{xcolor}
\usepackage[caption=false]{subfig}
\usepackage{amssymb}
\usepackage{amsmath}
\usepackage{commath}
\usepackage{verbatim}
\usepackage[normalem]{ulem}

\usepackage[colorlinks = true,
            linkcolor = red,
            urlcolor  = blue,
            citecolor = red,
            anchorcolor = blue]{hyperref}

\newcommand{\nue}{$\nu_e$\xspace}

\newcommand{\numu}{$\nu_{\mu}$\xspace}

\newcommand{\npsel}{1$e$N$p$0$\pi$\xspace}
\newcommand{\zpsel}{1$e$0$p$0$\pi$\xspace}

\newcommand{\dedx}{d$E$/d$x$\xspace}

\newcommand{\ke}{\ensuremath{\textrm{KE}\xspace}}
\newcommand{\npi}{\ensuremath{\pi^0}\xspace}

\newcommand{\Bern}{Universit{\"a}t Bern, Bern CH-3012, Switzerland}
\newcommand{\BNL}{Brookhaven National Laboratory (BNL), Upton, NY, 11973, USA}
\newcommand{\UCSB}{University of California, Santa Barbara, CA, 93106, USA}
\newcommand{\Cambridge}{University of Cambridge, Cambridge CB3 0HE, United Kingdom}
\newcommand{\CIEMAT}{Centro de Investigaciones Energ\'{e}ticas, Medioambientales y Tecnol\'{o}gicas (CIEMAT), Madrid E-28040, Spain}
\newcommand{\Chicago}{University of Chicago, Chicago, IL, 60637, USA}
\newcommand{\Cincinnati}{University of Cincinnati, Cincinnati, OH, 45221, USA}
\newcommand{\CSU}{Colorado State University, Fort Collins, CO, 80523, USA}
\newcommand{\Columbia}{Columbia University, New York, NY, 10027, USA}
\newcommand{\Edinburgh}{University of Edinburgh, Edinburgh EH9 3FD, United Kingdom}
\newcommand{\FNAL}{Fermi National Accelerator Laboratory (FNAL), Batavia, IL 60510, USA}
\newcommand{\Granada}{Universidad de Granada, Granada E-18071, Spain}
\newcommand{\Harvard}{Harvard University, Cambridge, MA 02138, USA}
\newcommand{\IIT}{Illinois Institute of Technology (IIT), Chicago, IL 60616, USA}
\newcommand{\KSU}{Kansas State University (KSU), Manhattan, KS, 66506, USA}
\newcommand{\Lancaster}{Lancaster University, Lancaster LA1 4YW, United Kingdom}
\newcommand{\LANL}{Los Alamos National Laboratory (LANL), Los Alamos, NM, 87545, USA}
\newcommand{\Louisiana}{Louisiana State University, Baton Rouge, LA, 70803, USA}
\newcommand{\Manchester}{The University of Manchester, Manchester M13 9PL, United Kingdom}
\newcommand{\MIT}{Massachusetts Institute of Technology (MIT), Cambridge, MA, 02139, USA}
\newcommand{\Michigan}{University of Michigan, Ann Arbor, MI, 48109, USA}
\newcommand{\Minnesota}{University of Minnesota, Minneapolis, MN, 55455, USA}
\newcommand{\NMSU}{New Mexico State University (NMSU), Las Cruces, NM, 88003, USA}
\newcommand{\Oxford}{University of Oxford, Oxford OX1 3RH, United Kingdom}
\newcommand{\Pitt}{University of Pittsburgh, Pittsburgh, PA, 15260, USA}
\newcommand{\Rutgers}{Rutgers University, Piscataway, NJ, 08854, USA}
\newcommand{\SLAC}{SLAC National Accelerator Laboratory, Menlo Park, CA, 94025, USA}
\newcommand{\SDSMT}{South Dakota School of Mines and Technology (SDSMT), Rapid City, SD, 57701, USA}
\newcommand{\Maine}{University of Southern Maine, Portland, ME, 04104, USA}
\newcommand{\Syracuse}{Syracuse University, Syracuse, NY, 13244, USA}
\newcommand{\TelAviv}{Tel Aviv University, Tel Aviv, Israel, 69978}
\newcommand{\Tennessee}{University of Tennessee, Knoxville, TN, 37996, USA}
\newcommand{\UTA}{University of Texas, Arlington, TX, 76019, USA}
\newcommand{\Tufts}{Tufts University, Medford, MA, 02155, USA}
\newcommand{\VTech}{Center for Neutrino Physics, Virginia Tech, Blacksburg, VA, 24061, USA}
\newcommand{\Warwick}{University of Warwick, Coventry CV4 7AL, United Kingdom}
\newcommand{\Yale}{Wright Laboratory, Department of Physics, Yale University, New Haven, CT, 06520, USA}

\begin{document}

\title{Differential cross section measurement of charged current \nue interactions without final-state pions in MicroBooNE}

\affiliation{\Bern}
\affiliation{\BNL}
\affiliation{\UCSB}
\affiliation{\Cambridge}
\affiliation{\CIEMAT}
\affiliation{\Chicago}
\affiliation{\Cincinnati}
\affiliation{\CSU}
\affiliation{\Columbia}
\affiliation{\Edinburgh}
\affiliation{\FNAL}
\affiliation{\Granada}
\affiliation{\Harvard}
\affiliation{\IIT}
\affiliation{\KSU}
\affiliation{\Lancaster}
\affiliation{\LANL}
\affiliation{\Louisiana}
\affiliation{\Manchester}
\affiliation{\MIT}
\affiliation{\Michigan}
\affiliation{\Minnesota}
\affiliation{\NMSU}
\affiliation{\Oxford}
\affiliation{\Pitt}
\affiliation{\Rutgers}
\affiliation{\SLAC}
\affiliation{\SDSMT}
\affiliation{\Maine}
\affiliation{\Syracuse}
\affiliation{\TelAviv}
\affiliation{\Tennessee}
\affiliation{\UTA}
\affiliation{\Tufts}
\affiliation{\VTech}
\affiliation{\Warwick}
\affiliation{\Yale}

\author{P.~Abratenko} \affiliation{\Tufts}
\author{J.~Anthony} \affiliation{\Cambridge}
\author{L.~Arellano} \affiliation{\Manchester}
\author{J.~Asaadi} \affiliation{\UTA}
\author{A.~Ashkenazi}\affiliation{\TelAviv}
\author{S.~Balasubramanian}\affiliation{\FNAL}
\author{B.~Baller} \affiliation{\FNAL}
\author{C.~Barnes} \affiliation{\Michigan}
\author{G.~Barr} \affiliation{\Oxford}
\author{J.~Barrow} \affiliation{\MIT}\affiliation{\TelAviv}
% ...

\collaboration{The MicroBooNE Collaboration}
\thanks{microboone\_info@fnal.gov}\noaffiliation

\date{\today}

\begin{abstract}
In this letter we present the first measurements of an exclusive electron neutrino cross section with the MicroBooNE experiment using data from the Booster Neutrino Beamline at Fermilab. These measurements are made for a selection of charged-current electron neutrinos without final-state pions. Differential cross sections are extracted in energy and angle with respect to the beam for the electron and the leading proton.  The differential cross section as a function of proton energy is measured using events with protons both above and below the visibility threshold. This is done by including a separate selection of electron neutrino events without reconstructed proton candidates in addition to those with proton candidates.
Results are compared to the predictions from several modern generators, and we find the data agrees well with these models.  The data shows best agreement, as quantified by $p$-value, with the generators that predict a lower overall cross section,  such as GENIE v3 and NuWro.
\end{abstract}

\addtolength{\tabcolsep}{5pt}


\maketitle

Many fundamental questions in neutrino physics are still unresolved ~\cite{NuReview} and will be addressed by upcoming experiments that use liquid argon detectors~\cite{DUNE:2020ypp,SBN}. These experiments will look for the appearance of electron neutrinos in a muon-neutrino beam to search for CP violation, measure the neutrino mass ordering, and explore longstanding anomalies. They will also address broader physics goals such as searching for dark matter particles in the beam, for which $\nu_{e}$ interactions are a dominant background, and characterizing supernova explosions, for which $\nu_{e}$ interactions are the primary signal. It is therefore vital to improve the modeling of $\nu_{e}$ interactions in argon to enable those searches with high sensitivity.

We present a measurement of \nue interactions in argon without final-state pions in MicroBooNE, both with and without visible protons. This analysis is the first \nue-argon cross section measurement in an exclusive final state and provides additional model discrimination relative to previous inclusive measurements. Also, as a first \nue cross section measurement on the Booster Neutrino Beamline (BNB)~\cite{BNB} at Fermilab, we provide a complementary result to previous measurements on argon~\cite{argoneutnue,ubnuminue,ubnuminueMCC9} performed on \nue events from the Neutrinos at the Main Injector (NuMI) beamline~\cite{numibeam}. This measurement also complements the differential electron neutrino cross-section measurement on a hydrocarbon target in a similar exclusive final state~\cite{MINERvA:2015jih}.

MicroBooNE has recently completed the first round of searches~\cite{eLEEPRL,PeLEE,DLPRD,WCPRD} for an excess of low-energy charged-current (CC) \nue interactions that could explain the MiniBooNE anomaly~\cite{MiniBooNE2020}, and did not observe an excess. The search for \nue events without visible final-state pions~\cite{PeLEE}, however, observed mild tension with the model used to predict the \nue interaction rate. Consistency was found to be at the 10\%--20\% level in terms of $p$-values after systematic uncertainties were constrained with a high-statistics measurement of CC \numu interactions from the same beam. 
In this letter we build on this result to perform a cross section measurement under the assumption of no new physics, with the goal of providing input to neutrino interaction model development.

The MicroBooNE detector~\cite{ubdetector} is a liquid argon time projection chamber (TPC).  The TPC is a 2.56~m by 2.32~m by 10.36~m volume filled with 85 metric tons of liquid argon.  As charged particles travel through the detector, they ionize the argon, and the ionization electrons drift in the applied electric field of 273 V/cm, to be detected by induction on two planes of wires and collected on the third plane of wires.  Each plane of wires has a different orientation (vertical, $+60^{\circ}$, $-60^{\circ}$) so that when they are read out in time they result in three different ``views'' that are combined to derive 3D images of neutrino interactions.
The detector also contains a light collection system, consisting of 32 photomultiplier tubes with fast timing resolution, that makes it possible to identify ionization electrons coincident with the neutrino beam arrival.

The neutrinos measured in this analysis come from the BNB.  They have an average energy of about 0.8 GeV and are primarily muon neutrinos, with only a 0.5\% contribution from electron neutrinos~\cite{mbflux}.  This analysis measures this intrinsic electron neutrino component using data collected from 2016--2018, corresponding to $6.86 \times 10^{20}$ protons on target (POT).

The neutrino flux simulation used in this analysis was developed by the MiniBooNE collaboration \cite{mbflux} and is modified to use the position of MicroBooNE.  Neutrino interactions in the detector argon are simulated using v3.0.6 G18\_10a\_02\_11a of the GENIE event generator~\cite{GENIEv3} with the MicroBooNE tune applied~\cite{ubtune}. 
There are several steps involved to simulate the detector response.  Particles are propagated through the detector using Geant4~\cite{geant4}, and then the charge and light produced by these particles is simulated with LArSoft~\cite{LArSoft}.  A simulation of the charge induced by drifting electrons is used for the wire and readout electronics response~\cite{ubsig1,ubsig2}.  Scintillation light propagation is modeled with a look-up table from a Geant4 simulation of photon propagation.  Data-driven electric field maps are used to take into account distortions in the electric field from space charge~\cite{ubefield,efieldcosmics}.  Ion recombination is simulated with a modified box model~\cite{argoneutrecomb}, and a time dependent simulation is used for the drift electron lifetime and wire response.  Cosmic rays are a significant background in MicroBooNE and are incorporated in a data-driven way by overlaying a simulated neutrino interaction onto cosmic data collected during periods of time when the neutrino beam was off.  This method also provides a data-driven incorporation of detector noise. 

Neutrino events are reconstructed in this analysis using the Pandora pattern-recognition toolkit~\cite{pandora}.  A set of algorithms first removes obvious cosmic-rays that cross the detector and then selects a neutrino candidate in time with the beam. Particles are reconstructed as showers or tracks within this neutrino candidate; typically electrons and photons are shower-like, while muons, charged pions, and protons are track-like.  The Pandora event reconstruction has been used for many previously published results by the MicroBooNE collaboration~\cite{gLEE,PeLEE,ubnuminue,ubnuminueMCC9,ubnumunp,ubccqexsec,ubccincl,ubccpi0,ubtrkmult,ubhportal,ubhsl}.  Additional tools are used on top of the Pandora pattern recognition to enhance shower-track separation, perform particle identification to separate proton and muon tracks~\cite{llrpid}, and to perform electron-photon separation for showers~\cite{PeLEE}.
Track and shower energies are measured separately.  Calorimetric energy reconstruction is performed for electromagnetic showers starting with the total energy clustered in the shower ($E_{\textrm{shr}}$).  This is corrected to account for inefficiencies in charge collection using a simulation of electrons and with this correction the reconstructed energy is defined as $E_{\textrm{reco}}$=$E_{\textrm{shr}}$/0.83. For tracks, the energy is estimated based on particle range~\cite{nist}.  Using simulation, the energy resolution is estimated to be 3\% for protons if their kinetic energy (KE) is greater than 50 MeV, and 12\% for electrons.  The absolute resolution on $\cos \theta$ is 0.01 for electrons and 0.03 for protons, where $\theta$ is the angle of the particle with respect to the beam.

We define true signal events as charged current  \nue interactions that contain an outgoing electron with $\textrm{KE}_{e} > 30$ MeV, and do not contain final-state charged pions with $\textrm{KE}_{\pi^{\pm}}>$ 40 MeV or any neutral pions.
Signal events are further characterized in terms of the leading proton kinetic energy.  Events with visible protons ($\ke_{p} \geq 50$ MeV) are defined as \npsel events.  Events without visible protons ($\ke_{p} < 50$ MeV), or events for which no proton exits the nucleus, are defined as \zpsel events~\cite{Suppl}. These \zpsel events are required to pass additional phase space restrictions on the electron energy ($E_e > 0.5$ GeV) and the angle between the neutrino beam and electron directions ($\cos\theta_{e}>0.6$).

We perform a differential cross section measurement in four kinematic variables: the electron energy, the electron angle with respect to the beam, the leading proton energy, and the leading proton angle with respect to the beam. 
All of these variables except the leading proton energy are measured for only the \npsel signal. The leading proton energy measurement includes both \zpsel and \npsel events with smearing allowed between these two samples. This is possible because \zpsel signal events by definition have a leading proton kinetic energy below 50 MeV, and therefore these events can be included as a single bin in the proton kinetic energy measurement from 0 to 50 MeV. This is the first measurement to characterize proton production in neutrino interactions across the visibility threshold.
Using the MicroBooNE tune of GENIE v3~\cite{ubtune}, \npsel events are predicted to be 60\% quasi-elastic (QE) neutrino interactions, 30\% meson exchange current (MEC), and with subdominant contributions from resonant (RES) (10\%) and deep inelastic scattering (DIS) (1\%) interactions; \zpsel events are mostly QE, with contributions from MEC and RES each at the 10\%--15\% level~\cite{Suppl}. The relative abundance of the different interaction types is not flat with respect to the measured variables which may provide some insight into the differences between models when data is compared to event generators.

Events are selected with separate criteria based on the presence or absence of candidate protons.  This selection strategy is the same as in Ref.~\cite{PeLEE}, although a few of the requirements have been updated to optimize the selections for a cross section measurement.  The main objective is to maintain sufficient \nue purity for a cross-section extraction while maximizing the \nue efficiency across the phase space of the measurement. For both the \npsel and \zpsel selections the largest increase in efficiency comes from a relaxed cut on the boosted decision trees (BDTs) used in the analysis.  These BDTs are the same, including the training, as those used in Ref.~\cite{PeLEE}. Additionally, for the \npsel selection, we relax the  requirements on proton vs muon particle identification, on the shower \dedx, and on the shower conversion distance. For the \zpsel selection we add requirements to increase the purity as needed for a cross-section measurement, particularly on the energy deposited per unit length (\dedx) at the start of the electron candidate, and by restricting the phase space to the highest-purity region with $\cos\theta_\textrm{e}^\textrm{reco}>0.6$ and E$_\textrm{e}^\textrm{reco} >$ 0.51 GeV. We find that with these selections an appropriate visibility threshold for the leading proton kinetic energy is 50 MeV, which is approximately where the \zpsel selection efficiency turns off and the \npsel efficiency turns on~\cite{Suppl}. Therefore, for \npsel selected events we also require that the leading reconstructed proton has $\ke_{p}^\textrm{reco} > 50$ MeV.

With the data sample used in this analysis, a total of 145.5 events are predicted in the \npsel selection, with a \npsel purity of 69\%.  We expect to select about 100 (2) true \npsel (\zpsel) events with an efficiency of 17\%.
 The largest backgrounds to the \npsel selection are events with final state \npi (\nue CC and \numu CC or NC interactions, for a total 15.3 predicted events), other \numu CC events (12.9 predicted events), and cosmic rays (6.8 predicted events). In the \zpsel selection about 10 (2) true \zpsel (\npsel) signal events are predicted with an efficiency of 12\% and \zpsel purity of 65\%; the total prediction is 17.6 events, and the largest background is from  interactions with final state $\pi^{0}$ mesons (2.8 predicted events). 

The prediction on the total number of selected events is subject to uncertainties from several sources. Variations in the flux prediction may come from uncertainties on the hadron production cross section and on the modeling of the beamline~\cite{mbflux,mbflux2}.  These are propagated to an uncertainty on the predicted event rate by reweighting the nominal simulation, and are found to be at the 6\% level and mostly flat in terms of the variables used in the analysis. 
Uncertainties on the neutrino interaction model are included based on the nominal tuned GENIE v3 simulation using a reweighting method for most of the sources and with a limited set of specific variations~\cite{ubtune}. 
The impact of the interaction model uncertainties is only evaluated on the efficiency and smearing for true signal events; the number of signal events is not varied as it is the quantity of interest for the cross-section measurement. These combine to a 4\% uncertainty on the total event prediction.
Uncertainties on the propagation of final state particles in the detector are assessed by varying re-interaction cross sections for charged pions and protons, again by reweighting~\cite{G4Reweight}.  These uncertainties are generally at the 1\% level, but grow to as high as 8\% at high proton energies. Uncertainties on detector modeling are assessed using dedicated samples that are produced by varying parameters related to specific detector effects to amounts compatible with estimates from MicroBooNE data.  These include space-charge effects, electron-ion recombination, light measurement, and wire response~\cite{DetSyst}. Overall, these effects combine to approximately a 5\% effect but can grow to 10\%--20\% at high electron and proton energies as well as for the \zpsel selection. Other subdominant uncertainties are due to the size of simulated samples, the POT measurement, and the estimate of the total number of argon nuclei in the detector. 

Covariance matrix formalism is used to include systematic uncertainties in the analysis, where the total systematic uncertainty covariance matrix $C^{\textrm{Syst}}$ is defined as the sum of the covariance matrices of each uncertainty (flux, cross section, re-interaction, detector, Monte Carlo statistics, POT, and the number of nuclei), with individual entries written as $C_{ij}$:
\begin{equation}
    C_{ij} = \frac{1}{N}\sum_{k=1}^{N} \left(n_i^{k} - n_i^{\textrm{CV}}\right)\left(n_j^{k} - n_j^\textrm{CV}\right).
\end{equation}
Here the covariance matrix is written in terms of bin indices $i$ and $j$, and constructed as a sum over systematic variations $k$ up until the total number of systematic variations $N$, with the central value bin content defined as $n^{\textrm{CV}}$ and the content of bin $i$ in variation $k$ defined as $n^{k}_{i}$.
Finally, statistical uncertainties from the data measurement are included as 
\begin{equation}\label{eq:totcov}
        C^\textrm{Tot} = C^\textrm{Syst} + C^\textrm{DataStat}, \\
\end{equation}
where $C^\textrm{DataStat}$ is diagonal with elements corresponding to the Poisson variance in each bin. Statistical uncertainties in the data are the leading source of uncertainty in this measurement.

\begin{figure*}
\subfloat[\label{sfig:ea_reco}]{
    \includegraphics[width=.49\linewidth]{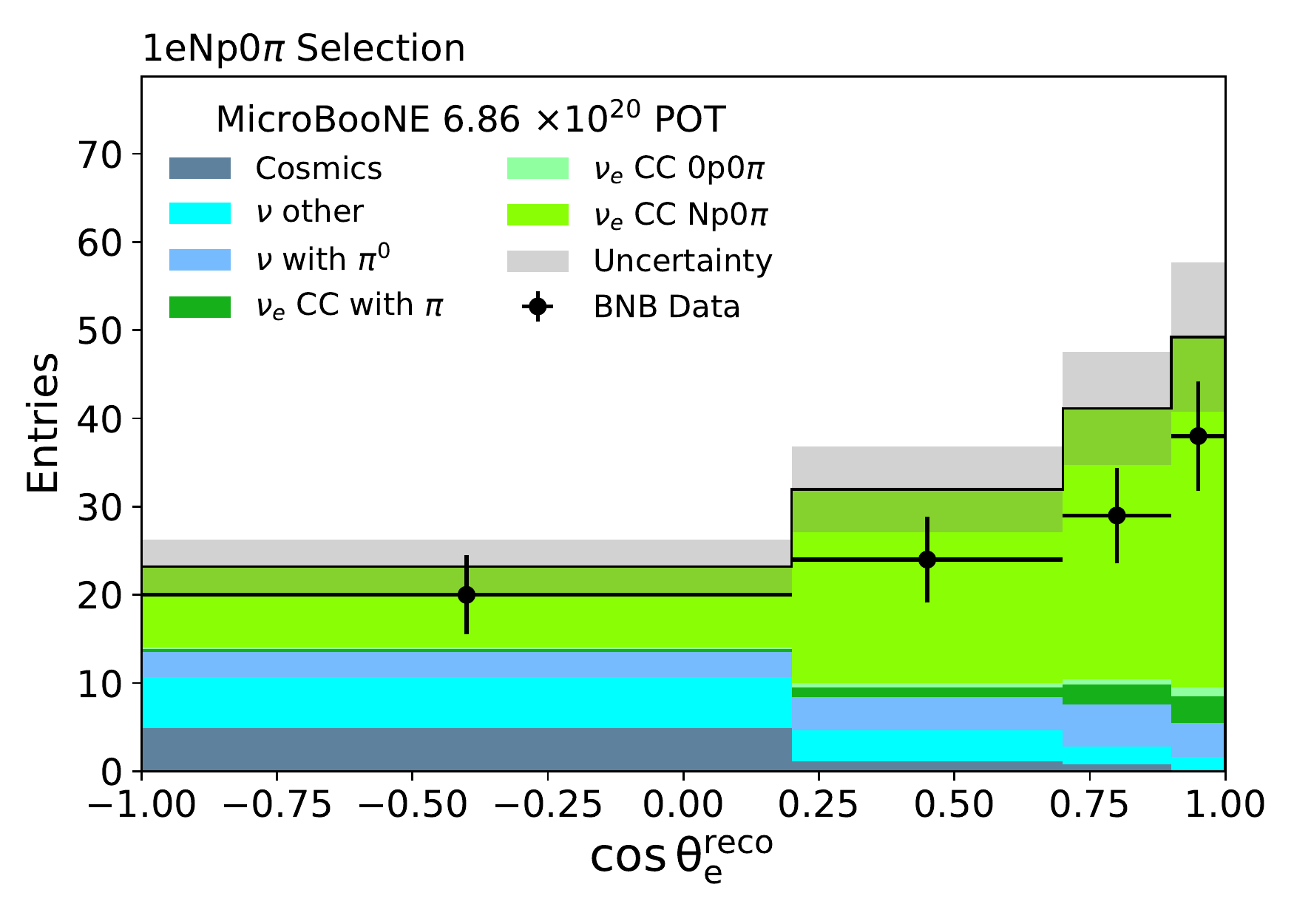}
}\hfill
\subfloat[\label{sfig:ee_reco}]{%
  \includegraphics[width=.49\linewidth]{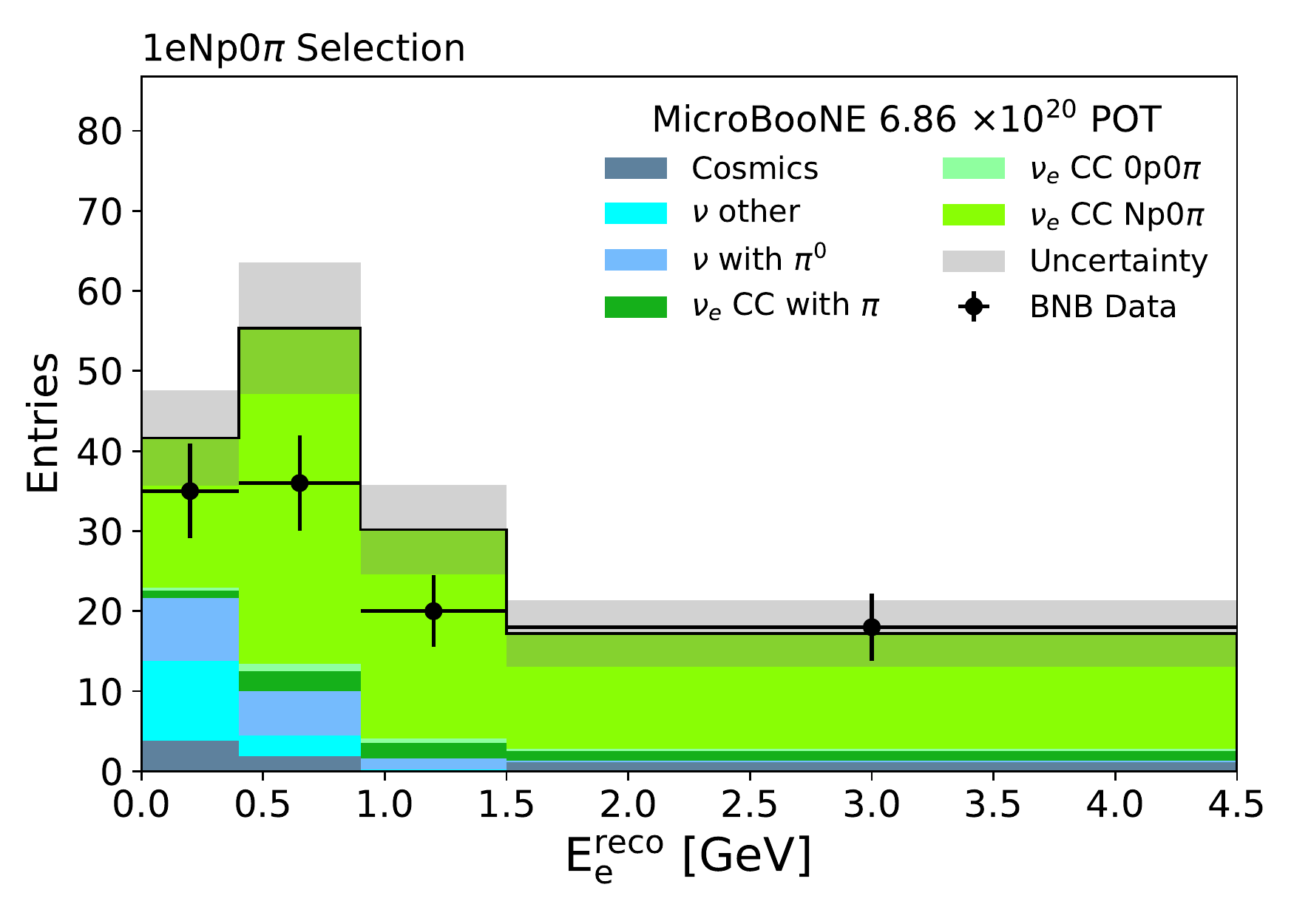}
}\\
\subfloat[\label{sfig:pa_reco}]{
    \includegraphics[width=.49\linewidth]{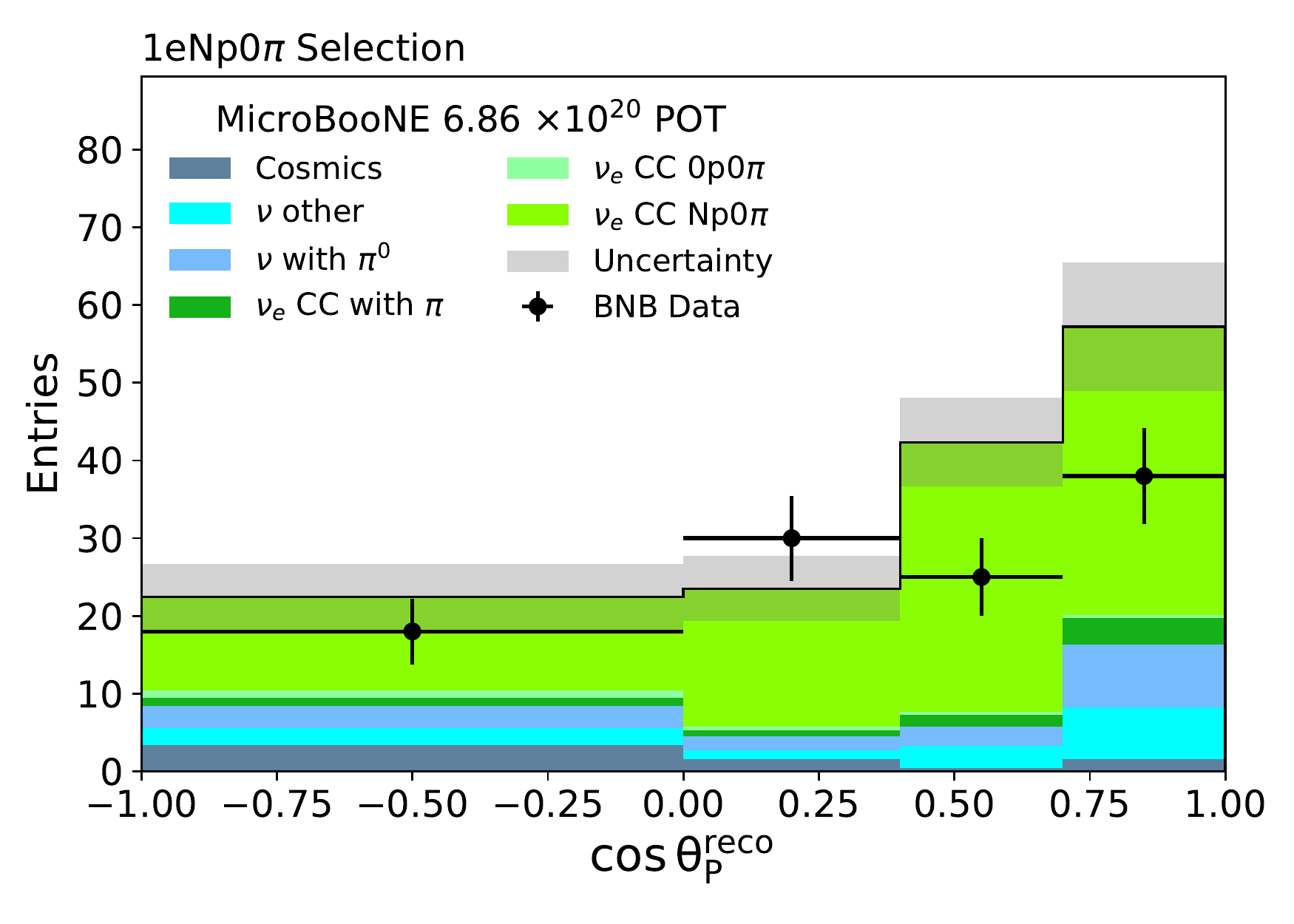}
}\hfill
\subfloat[\label{sfig:pe_reco}]{%
  \includegraphics[width=.49\linewidth]{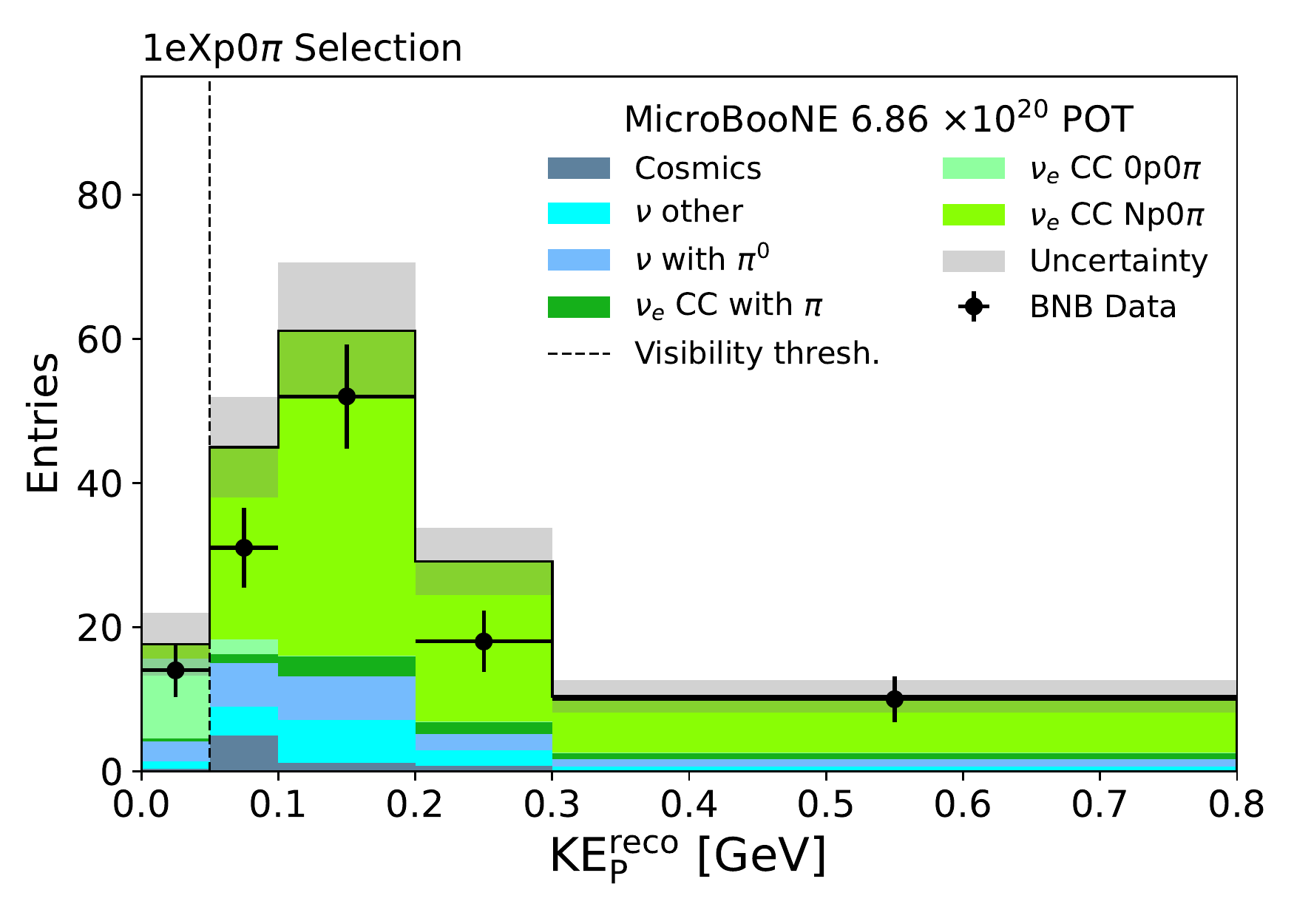}
}
\caption{\label{fig:obs} The observed number of events in data compared to the simulated prediction using the MicroBooNE tune of GENIE v3. The selection used is reported in each panel.
The \npsel selection is used for \protect\subref{sfig:ea_reco} the angle between the neutrino beam and electron direction, \protect\subref{sfig:ee_reco} the electron energy, and \protect\subref{sfig:pa_reco} the angle between the neutrino beam and leading proton direction.  The 1$e$X$p$0$\pi$ = (\zpsel OR \npsel) selection is used for \protect\subref{sfig:pe_reco}, the leading proton kinetic energy, where events selected with the \zpsel selection populate the leftmost bin and events from the \npsel selection populate the other bins.}
\end{figure*}

The observed distributions for the four variables considered in this analysis are shown in Fig.~\ref{fig:obs}, where the data is overlaid on top of the nominal simulation based on the tuned version of GENIE v3~\cite{ubtune}. The data sample consists of 111 events selected with the \npsel selection and an additional 14 events with the \zpsel selection. The simulation predicts more events than the data, especially at forward angles with respect to the beam and at intermediate energies. These are similar observations to those presented in Ref.~\cite{PeLEE}.

To extract the cross section from the observed number of events we first define a response matrix, which maps the generated signal events in the true variable space to the observed signal events after selection in the reconstructed space. The off-diagonal elements of the response matrix define the amount of smearing between true and reconstructed bins. Both \zpsel and \npsel events are included in the response matrix for the proton energy, with \zpsel events in a single bin and \npsel events in the other bins.  This means that smearing is included between these selections through the off-diagonal elements.
The other variables use only \npsel events.
Due to the limited size of the selected data sample the bin width is typically larger than the resolution on the measured variables so smearing is limited and most events fall into the correct bins with $>$70\% across all variables and $>$90\% for electron angle.
An unfolded differential cross-section measurement in the true-space bin $i$ for the variable $x$ measured in reconstructed-space bin $j$ is defined as:
\begin{equation}\label{eq:xsec}
    \left \langle\frac{\textrm{d}\sigma}{\textrm{d}x}\right \rangle_i = \frac{\sum_j U_{ij}(n_j - b_j)}{N_{\text{target}} \times \phi \times (\Delta x)_i},
\end{equation}
where $U$ is the unfolding matrix, $n$ is the number of data events, $b$ is the number of background events, $N_{\textrm{target}}$ is the number of nucleons, $\phi$ is the integrated electron neutrino flux, and $(\Delta x)_i$ is the measured bin width in the variable $x$. The unfolding matrix $U$ is used in place of the inverse of the response matrix $R^{-1}$ to avoid instabilities in the cross-section result from a direct matrix inversion.
We extract the cross section using an unfolding procedure based on the D'Agostini method~\cite{DAgUnfold} with three iterations. 
This number of iterations is found to give results that are stable and with limited bin-to-bin fluctuations.
In the cross-section extraction, we use a number of nucleons equal to $4.3912 \times 10^{31}$, and a POT-integrated BNB \nue flux of $2.73 \times 10^{9}$ cm$^{-2}$, which is taken to be the reference flux~\cite{Koch:2020oyf} of the measurement and used as a constant value. As described in a previous MicroBooNE publication~\cite{MicroBooNECollaboration:2021enn}, this method allows for a consistent treatment of flux uncertainties. The uncertainties on the total prediction (Eq.~\ref{eq:totcov}) are analytically propagated through the unfolding procedure to obtain a covariance matrix in unfolded cross section~\cite{Bourbeau:2018wiw}.

\begin{figure*}
\subfloat[\label{sfig:ea}]{
    \includegraphics[width=.49\linewidth]{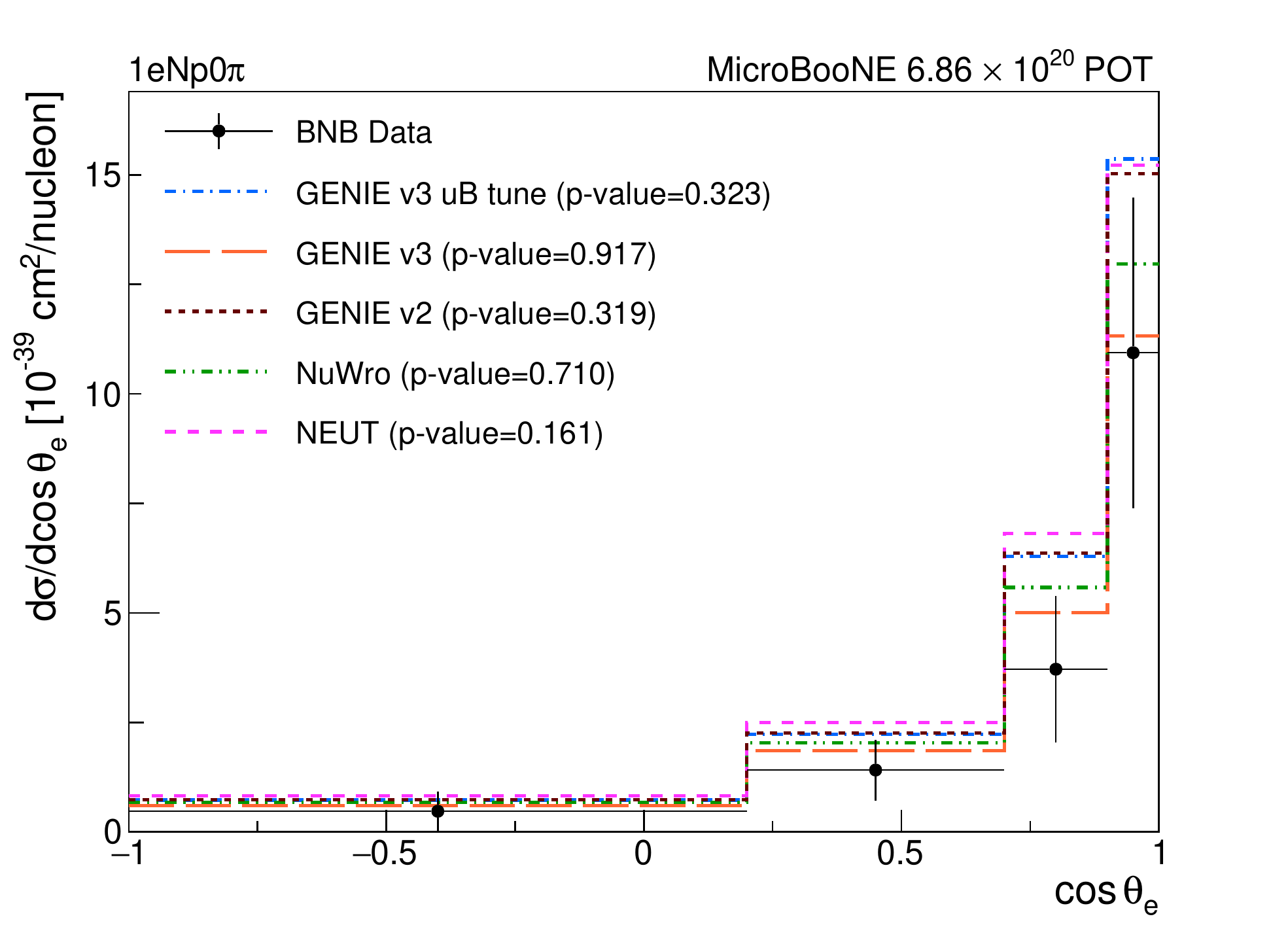}
}\hfill
\subfloat[\label{sfig:ee}]{%
  \includegraphics[width=.49\linewidth]{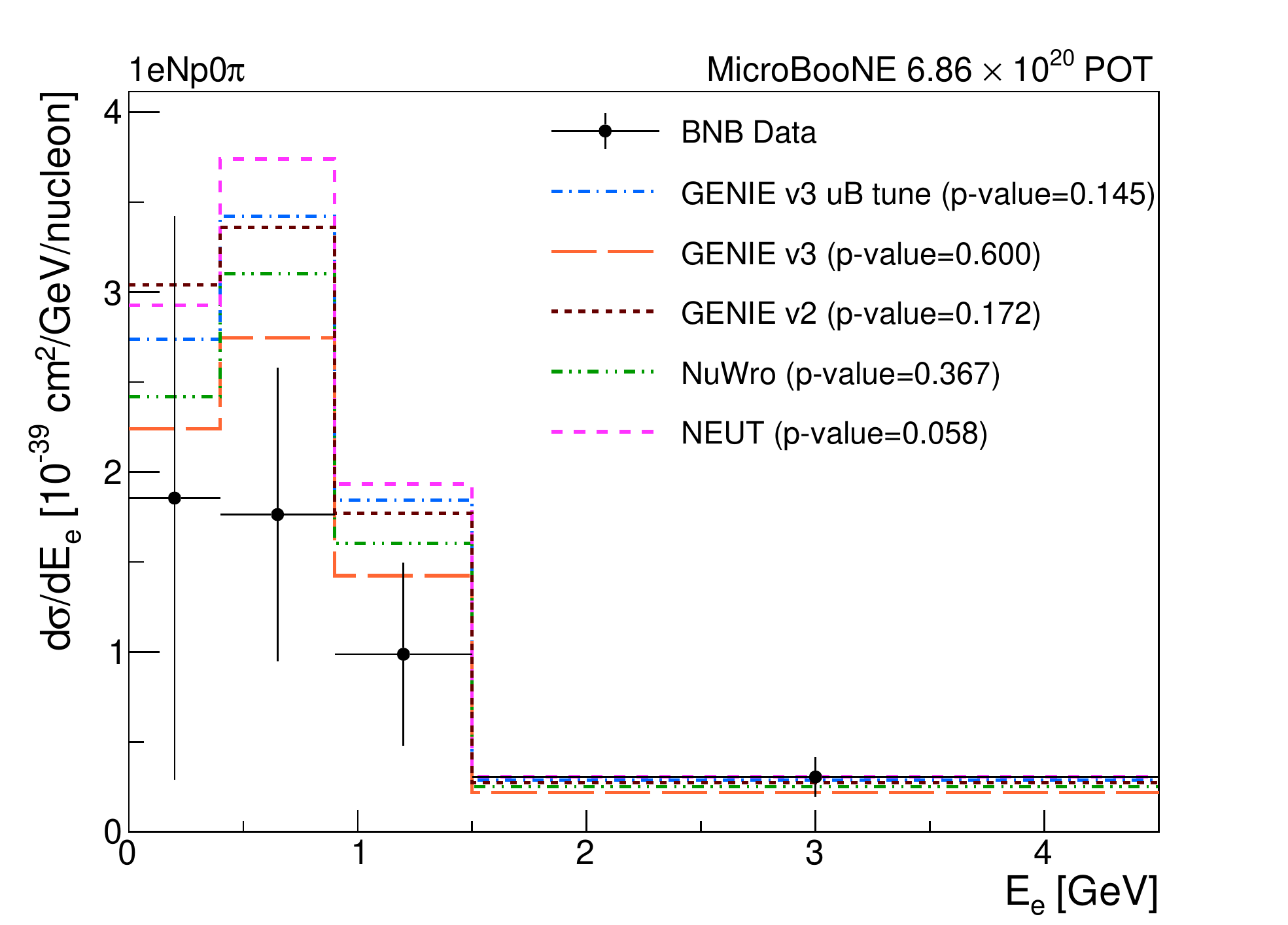}
}\\
\subfloat[\label{sfig:pa}]{
    \includegraphics[width=.49\linewidth]{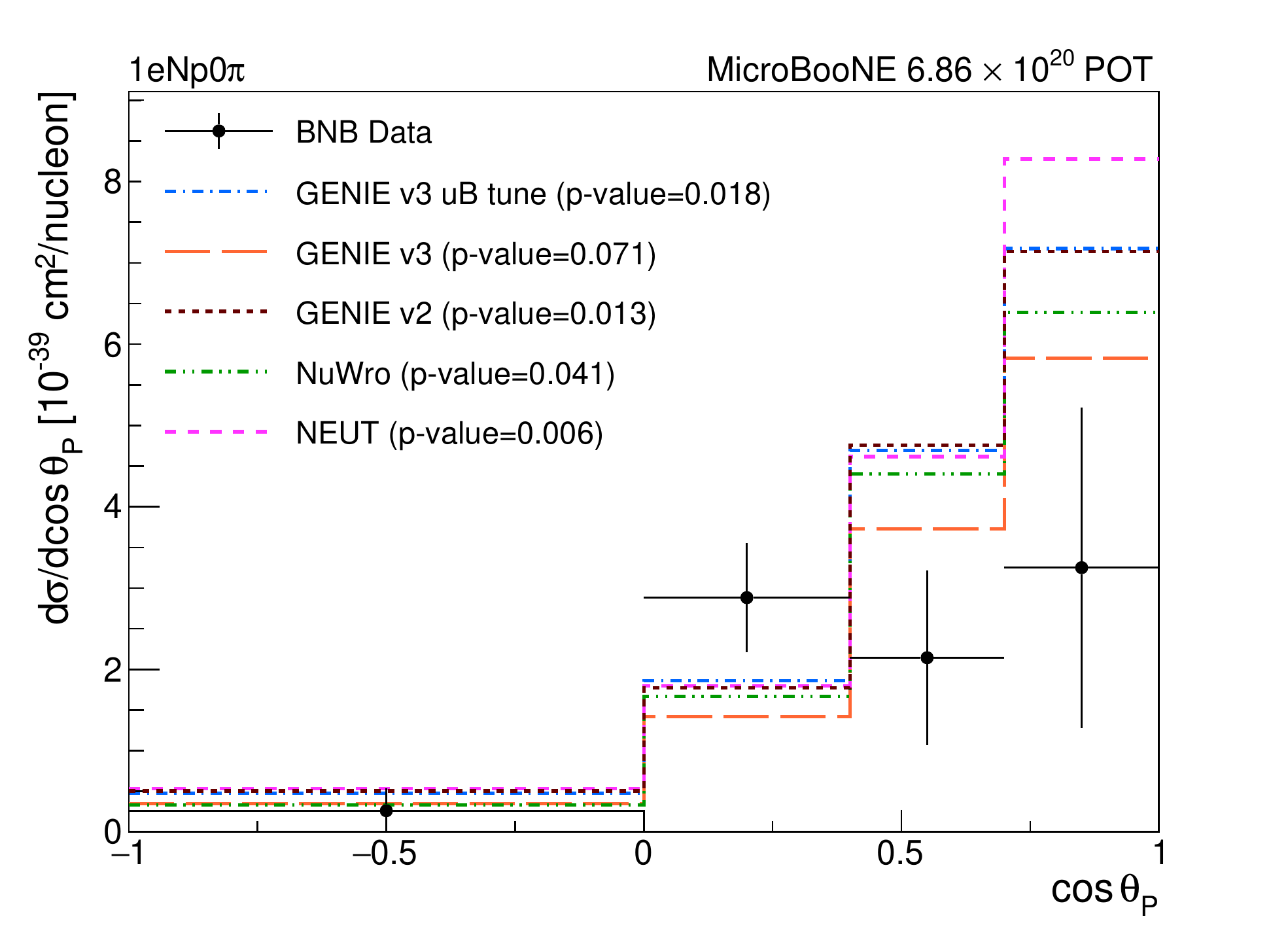}
}\hfill
\subfloat[\label{sfig:pe}]{%
  \includegraphics[width=.49\linewidth]{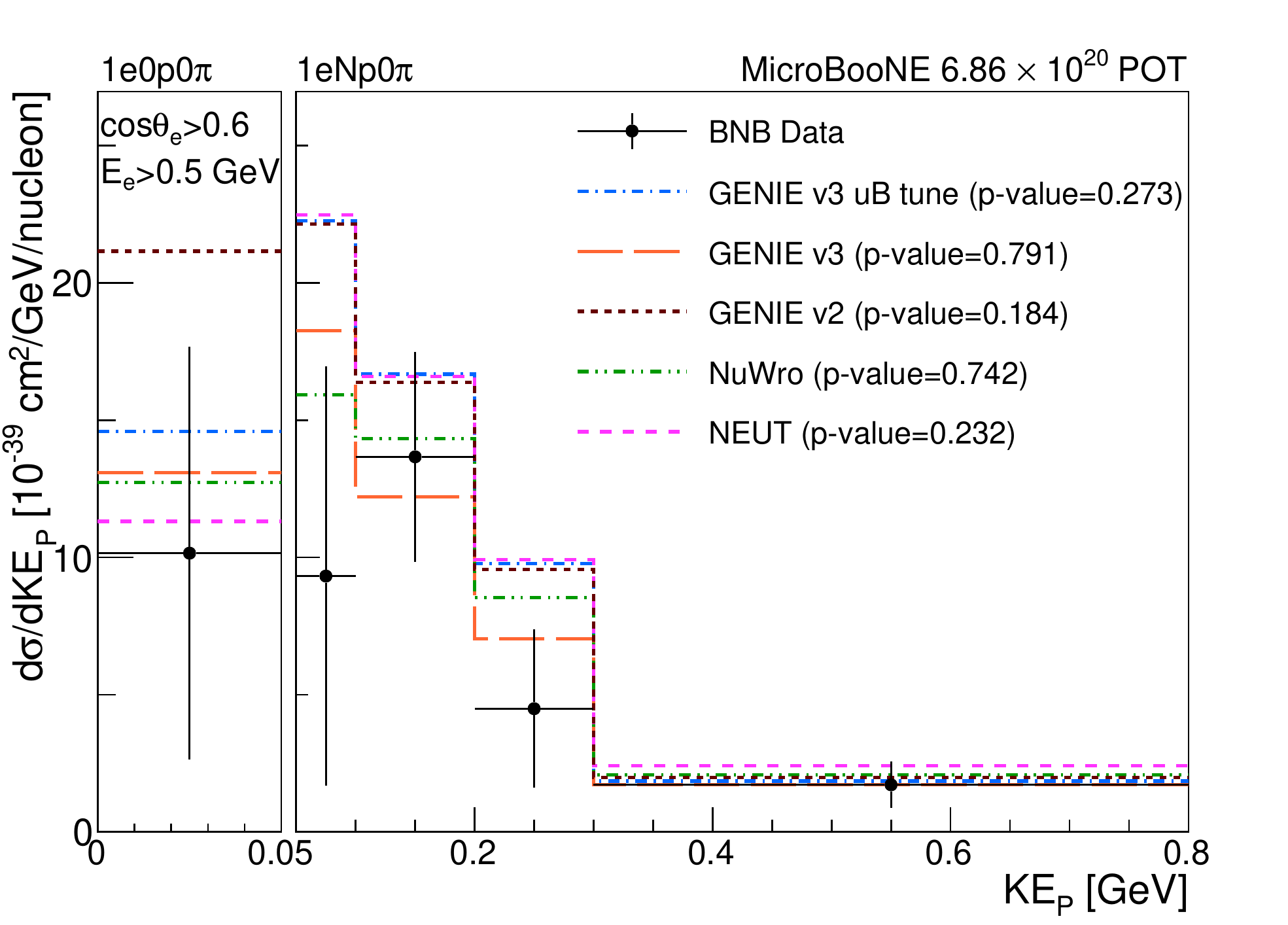}
}
\caption{\label{fig:xsec} Differential cross sections from unfolded data and comparisons with predictions from different generators. The signal definition is reported for each panel: \npsel is used for \protect\subref{sfig:ea} the angle between the neutrino beam and electron direction, \protect\subref{sfig:ee} the electron energy, \protect\subref{sfig:pa} the angle between the neutrino beam and the leading proton direction, and the right panel of \protect\subref{sfig:pe} the leading proton kinetic energy.  An additional phase space restriction is applied to the leftmost panel of \protect\subref{sfig:pe}.  Compatibility is evaluated in terms of $p$-values, and reported in the legends.}
\end{figure*}

The resulting cross sections are presented in Fig.~\ref{fig:xsec}, where they are compared to a number of modern generators: the MicroBooNE tune of GENIE v3.0.6~\cite{ubtune}, GENIE v3.0.6 G18\_10a\_02\_11a~\cite{GENIEv3}, GENIE v2.12.2~\cite{Geniev2,Geniev2um}, NuWro 19.02.1~\cite{NuWro1,NuWro2}, and NEUT v5.4.0~\cite{Hayato:2009zz,NEUT}. 
These generators have different initial state nuclear models (GENIE v2 uses a relativistic Fermi gas, while the others use a local Fermi gas), quasi-elastic models (GENIE v3 and NEUT use Valencia~\cite{Nieves:2011pp, Schwehr:2016pvn, Gran:2013kda}, GENIE v2 and NuWro use Llewellyn Smith~\cite{LlewellynSmith:1971uhs}), and MEC models (GENIE v2 uses an empirical model, and the others the Valencia model).  Details about the models used in these generators and a more complete description of their differences are found in other MicroBooNE publications~\cite{ubccqexsec,ubnuminueMCC9,ubnumunp} and a summary table presented in~\cite{NOvAccmu}.
We assess the agreement with these generators by computing $\chi^2$ values and the $p$-values corresponding to the upper tail of the cumulative distribution for the $\chi^2$ per degrees of freedom.

\begin{table}
 \caption{\label{tab:pvals} Agreement between unfolded data and generator neutrino interaction models represented as $p$-values. }
\begin{tabular}{l c c c c} 
 \hline \hline
  Generator & $\cos \theta_{e}$ & $\textrm{E}_{e}$ & $\cos \theta_{p}$ & $\textrm{KE}_{p}$\\ \hline 
 GENIE v3 uB tune & 0.323 & 0.145 & 0.018 & 0.273\\ 
 GENIE v3 & 0.917 & 0.600 & 0.071 & 0.791\\ 
 GENIE v2 & 0.319 & 0.172 & 0.013 & 0.184\\ 
NuWro & 0.710 & 0.367 & 0.041 & 0.742\\ 
NEUT & 0.161 & 0.058 & 0.006 & 0.232\\ 
 \hline \hline
\end{tabular}
\end{table}

While all generators are in reasonable agreement with the data, the level of agreement differs depending on the generator and the variable as shown in Table~\ref{tab:pvals}.
The data indicate a preference for GENIE v3 and NuWro, both of which have a smaller overall electron neutrino prediction.
Compared to the default GENIE v3, the MicroBooNE tune enhances the QE and MEC components and tends to over-predict, especially at intermediate energies. The lowest $p$-values are obtained for NEUT, which predicts the largest overall cross section, especially at forward proton angles, and GENIE v2, which has the largest prediction for \zpsel events, partly due to its empirical MEC model~\cite{Katori:2013eoa} with no Pauli blocking. 
The discrepancy between data and generator models is largest in leading proton angle, with $p$-values that range from 1\% to 7\%, and is most pronounced in the forward direction.  Future measurements with more statistics will be able to further explore these features.
More information about these results is provided in supplementary material, including tabulated cross-section values, $\chi^2$ values, the background-subtracted observations, covariance matrices, and response matrices~\cite{Suppl}.

In summary, this letter presents the first differential \nue-argon cross-section measurement without pions in the final state in electron angle and energy as well as leading proton angle and energy, where the proton energy is characterized both above and below the visibility threshold. 
The findings are typically in agreement with predictions from modern generators, except for tension in the proton angle, with an overall preference for those with lower total cross section.
These results provide input for further tuning of generators towards an improved \nue prediction for future new-physics searches in MicroBooNE, SBN~\cite{SBN}, and DUNE~\cite{DUNE:2020fgq}. While this result is statistically limited, an approximately equivalent data set from later run periods remains to be analyzed and can be used, in addition to possible reconstruction and selection improvements, for future cross-section measurements.

This document was prepared by the MicroBooNE collaboration using the resources of the Fermi National Accelerator Laboratory (Fermilab), a U.S. Department of Energy, Office of Science, HEP User Facility. Fermilab is managed by Fermi Research Alliance, LLC (FRA), acting under Contract No. DE-AC02-07CH11359.  MicroBooNE is supported by the following: the U.S. Department of Energy, Office of Science, Offices of High Energy Physics and Nuclear Physics; the U.S. National Science Foundation; the Swiss National Science Foundation; the Science and Technology Facilities Council (STFC), part of the United Kingdom Research and Innovation; the Royal Society (United Kingdom); and the UK Research and Innovation (UKRI) Future Leaders Fellowship. Additional support for the laser calibration system and cosmic ray tagger was provided by the Albert Einstein Center for Fundamental Physics, Bern, Switzerland. We also acknowledge the contributions of technical and scientific staff to the design, construction, and operation of the MicroBooNE detector as well as the contributions of past collaborators to the development of MicroBooNE analyses, without whom this work would not have been possible. For the purpose of open access, the authors have applied a Creative Commons Attribution (CC BY) public copyright license to any Author Accepted Manuscript version arising from this submission.

\bibliography{biblio}

\clearpage

\end{document}